\documentclass[epj]{svjour}

\usepackage{graphicx}
\usepackage{fancyhdr}
\usepackage{amssymb}
\usepackage{epsfig}
\usepackage{graphicx}
\usepackage{cite}
\usepackage{multirow}
\usepackage{longtable}
\usepackage{lscape}
\usepackage{bbm}
\def\makeheadbox{{%  remove EPJC header
 }}
\def\mytitle{My title} 
\def\myauthors{My name}  
\def\mytype{My type of session}
\def\mysession{My session}

\newcommand{\bi}{\begin{itemize}}
\newcommand{\ei}{\end{itemize}}

\newcommand{\be}{\begin{equation}}
\newcommand{\ee}{\end{equation}}
\newcommand{\bea}{\begin{eqnarray}}
\newcommand{\eea}{\end{eqnarray}}

\newcommand{\stheta}{\sin^2 2 \theta_{13}}

\newcommand{\ie}{{\it i.e.}}

\newcommand{\eg}{{\it e.g.}}

\newcommand{\cf}{{\it cf.}}

\newcommand{\etc}{{\it etc.}}
\newcommand{\eq}{Eq.}

\newcommand{\fig}{Fig.}

\newcommand{\Sec}{Sec.}

\newcommand{\Tab}{Table}

\newcommand{\equ}[1]{\eq~(\ref{equ:#1})}
\newcommand{\figu}[1]{\fig~\ref{fig:#1}}

\def\mytitle{Constructing Textures in Extended Quark-Lepton Complementarity}
\def\myauthors{Florian~Plentinger}
\def\mytype{Contributed Talk}
\def\mysession{Flavor Physics}

\begin{document}
\title{Constructing Textures in Extended Quark-Lepton Complementarity}
\subtitle{}
\author{Florian~Plentinger
\thanks{\emph{Email:} florian.plentinger@physik.uni-wuerzburg.de}%
}
\institute{
{
	Institut f{\"u}r Theoretische Physik und Astrophysik \\
	Universit{\"a}t W{\"u}rzburg \\
	D-97074 W{\"u}rzburg, Germany
}
}

\date{}

\abstract{
We systematically construct an extensive list of realistic mass matrix textures for leptons. For this set of matrices, we discuss the parameter space, and illustrate how these textures could be generated in explicit models from flavor symmetries.
}

\maketitle

\section{Introduction}
\label{sec:introduction}
Numerous theories have been considered to understand the striking difference between quark and lepton mixings and their mass hierarchies. A promising approach is the introduction of flavor symmetries. An example are spontaneously broken flavor symmetries generating masses from higher-dimension terms via the Froggatt-Nielsen mechanism \cite{Froggatt:1978nt}. In such approaches, effective dimension-$n$ mass terms, proportional to $\epsilon^n$, arise, where $\epsilon$ depends on the flavon vacuum expectation value (VEV) suppressed by the mass of superheavy fermions. In this way, mass matrix textures with $\epsilon$-powers as entries are obtained. Consequently, such a matrix structure contains information on the hierarchy among matrix elements and goes beyond approaches, \eg, using texture zeros.
\\
\indent However, many flavor symmetries and scenarios of mass generation are possible. Therefore, we suggest in \cite{Plentinger:2006nb,Plentinger:2007px} a systematic bottom-up approach. Thereby, we do not start with a symmetry to generate textures. Instead, we systematically construct a comprehensive list of mass matrix textures from very generic assumptions, for effective lepton masses and for the seesaw mechanism \cite{seesaw}. In our approach, we parameterize all mass ratios in powers of a small quantity $\epsilon \simeq 0.2$, which is of the order of the Cabibbo angle $\theta_C$, and allow the mixing angles to be either $\pi/4$ or use powers of $\epsilon$. This is the hypothesis of extended quark-lepton complementarity \cite{Plentinger:2006nb,Plentinger:2007px} and is motivated by the usual quark-lepton complementarity \cite{qlc}, various models with flavor symmetries predicting maximal mixing and the Froggatt-Nielsen mechanism.
\\
\indent From our resulting list of mass mass matrices, we investigate the parameter space, \eg, with respect to mixing angle and mass hierarchy distributions and special cases. Furthermore, we demonstrate how these textures could be generated in explicit models from flavor symmetries.

\section{Effective Mass Matrices}
\label{sec:SM}
\begin{figure*}[t!]
\begin{center}
\includegraphics[width=0.9\textwidth]{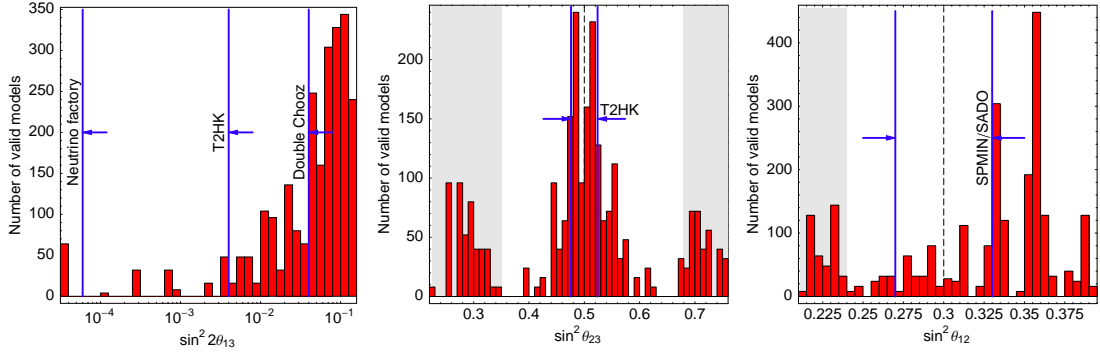}
\end{center}
\caption{\label{fig:histeps02} Distributions of $\stheta$ (left), $\sin^2 \theta_{23}$ (middle), and $\sin^2 \theta_{12}$ (right) of Yukawa coupling matrices for $\epsilon=0.2$ (figure taken from \cite{Plentinger:2006nb}). The bars show the number of selected Yukawa coupling matrices per bin, \ie, per specific parameter range. The gray-shaded regions mark the current $3 \sigma$-excluded regions.}
\end{figure*}
\begin{table*}
 \begin{center}
\begin{tabular}{|c||c|c||c|c|}\hline&&&&\vspace{-3mm}\\
 \# & $M_\ell$ & $M_\nu^\mathrm{ Maj}$ & \begin{minipage}{2.3cm}\centering
$(\theta^\ell_{12}, \theta^\ell_{13}, \theta^\ell_{23})$\\ $(\theta^\nu_{12}, \theta^\nu_{13}, \theta^\nu_{23})$\\ $(\delta^\ell,\delta^\nu, \widehat{\varphi}_1, \widehat{\varphi}_2)$ \end{minipage} & {$(\theta_{12}, \theta_{13}, \theta_{23})$} \\&&&&\vspace{-3mm}\\ \hline&&&&\vspace{-2.8mm}\\
17 & $\left(
\begin{array}{ccc}
 0 & \epsilon ^2 & \epsilon  \\
 0 & \epsilon ^2 & 1 \\
 0 & \epsilon ^2 & 1
\end{array}
\right)$  & 
 $\left(
\begin{array}{ccc}
 1 & \epsilon ^2 & 1 \\
 \epsilon ^2 & \epsilon  & \epsilon ^2 \\
 1 & \epsilon ^2 & 1
\end{array}
\right)$ & \begin{minipage}{2.3cm}\begin{center}
 $(\frac{\pi}{4},\, \epsilon,\, \frac{\pi}{4})$ \vspace{.1cm}\\
 $(\epsilon,\, \frac{\pi}{4},\, 0)$ \vspace{.1cm}\\
 $(\pi,\, \xi,\,\pi,\,\xi+\pi)$
\end{center}\end{minipage} & 
$\left(35.2^\circ, 3.8^\circ, 50.8^\circ\right)$
\\&&&&\vspace{-3mm}\\ \hline &&&&\vspace{-2.8mm}\\
18 & $\left(
\begin{array}{ccc}
 0 & \epsilon ^2 & \epsilon  \\
 0 & \epsilon ^2 & 1 \\
 0 & \epsilon ^2 & 1
\end{array}
\right)$  & 
 $\left(
\begin{array}{ccc}
 1 & \epsilon ^2 & 1 \\
 \epsilon ^2 & \epsilon  & \epsilon ^2 \\
 1 & \epsilon ^2 & 1
\end{array}
\right)$ & \begin{minipage}{2.3cm}\begin{center}
 $(\frac{\pi}{4},\, \epsilon,\, \frac{\pi}{4})$ \vspace{.1cm}\\
 $(\epsilon,\, \frac{\pi}{4},\, \epsilon^2)$ \vspace{.1cm}\\
 $(\pi,\, \pi,\,\pi,\,0)$
\end{center}\end{minipage} & 
$\left(33.6^\circ, 3.1^\circ, 52.2^\circ\right)$\\[5.3mm] \hline
 \end{tabular}
 \end{center}
\caption{Selected examples of mass matrix textures from \cite{Plentinger:2006nb}. For the mass matrices of charged leptons we use the basis in which the mixings of the right-handed fields are zero. Shown are also the corresponding mixing angles and phases of $U_\ell, U_\nu$ (with $\xi\in\{0,\,\pi\}$), as well as the mixing angles of $U_\mathrm{ PMNS}$.}
\label{tab:textures}
\end{table*}
In the Standard Model with massive Majorana neutrinos, the effective lepton mass terms have the form
\begin{equation}\label{equ:massterms}
 \mathcal{L}_\mathrm{ M}=-(M_\ell)_{ij}e_ie^c_j-\frac{1}{2}(M_\nu^\mathrm{ Maj})_{ij}\nu_i\nu_j+\mathrm{h.c.},
\end{equation}
where $e_i$ and $\nu_i$ are the left-handed charged leptons and neutrinos, $e_i^c$ the right-handed charged leptons, and $i=1,2,3$ is the generation index. The mass ratios of charged leptons and normal hierarchical (NH) neutrinos\footnote{In this talk we restrict ourselves to NH. However, in \cite{Plentinger:2006nb,Plentinger:2007px} we also discuss scenarios of inverted hierarchical and degenerate neutrinos.} are roughly given as
\be
\label{equ:massratios}
\begin{array}{lll}
m_e:m_\mu:m_\tau&=&\epsilon^4:\epsilon^2:1,\\
m_1:m_2:m_3&=&\epsilon^2:\epsilon:1.
\end{array}
\ee
The best-fit values of the mixing angles of the lepton mixing matrix (\cf, tribimaximal mixing \cite{tbm})
\be
\label{equ:PMNS}
U_\mathrm{PMNS}=U_\ell^\dagger U_\nu=\widehat{U}_\ell^\dagger D_\nu \widehat{U}_\nu K_\nu
\ee
are
\be
\label{equ:PMNSmixings}
\theta_{12}=\pi/4-\epsilon,\quad \theta_{13}=0,\quad \theta_{23}=\pi/4,
\ee
where $U_\mathrm{ PMNS}$ is  on the standard form, $U_\ell$ and $U_\nu$ are the charged lepton and neutrino mixing matrices respectively, $\widehat{U}_\ell$ and $\widehat{U}_\nu$ are CKM-like matrices, $D_\nu=\mathrm{diag}(1,e^{i\widehat{\varphi}_1},e^{i\widehat{\varphi}_2})$, and $K_\nu=\mathrm{diag}(e^{i\widehat{\phi}_1},e^{i\widehat{\phi}_2},1)$. Note that we have removed unphysical phases and have used the parameterization of a general unitary $3\times3$ matrix
\begin{equation}
\label{equ:unitary}
 U^\mathrm{ unitary}=D\widehat{U}K.
\end{equation}
Here, $D$ and $K$ are diagonal phase matrices, \ie, $D=\mathrm{diag}\left(e^{{i}\varphi_{1}},e^{{i}\varphi_{2}},e^{{i}\varphi_{3}}\right)$, $K=\mathrm{diag}\left(e^{{i}\alpha_{1}},e^{{i} \alpha_{2}}, 1 \right)$, and $\widehat{U}$ is a CKM-like matrix. The mass matrices are diagonalized by
\begin{equation}
 \label{equ:lepdiag}
M_\ell=U_\ell\,M_\ell^\mathrm{ diag}\,U_{\ell'}^\dagger,\quad  M_\nu^\mathrm{ Maj}=U_\nu M_\nu^\mathrm{ diag}U_\nu^T,
\end{equation}
where the mass matrices, in the mass basis, are $M_\ell^\mathrm{ diag}=m_\ell\,\mathrm{diag}(\epsilon^4,\epsilon^2,1)$ and $M_\nu^\mathrm{ diag}=m_\nu\,\mathrm{diag}(\epsilon^2,\epsilon,1)$.
\\
\indent From \equ{PMNS}, one can see that more than one combination of $U_\ell$ and $U_\nu$ can lead to the same PMNS mixing matrix. This ambiguity could be circumvented, \eg, by rotating to the basis where the charged leptons are diagonal, or by using invariants. However, we should keep in mind the origin of $U_\ell$ and $U_\nu$, \ie, that the mass matrices might be generated by a flavor symmetry. Therefore, we do not use such a simplification in our model-independent bottom-up approach in order not to loose or conceal this information. Later, we will identify the origin of our mass matrix textures by explicit models from flavor symmetries.
\\
\indent In order to obtain a comprehensive set of realistic mass matrices, we extend the parameterization described above and use the hypothesis of extended quark-lepton complementarity, \ie, all masses and mixings are powers of a small quantity $\epsilon\simeq\theta_C$. For the mixings, $\epsilon^0$ is interpreted as $\pi/4$. Therefore, we generate all possible combinations of \equ{PMNS} with $\theta_{ij}^x\in\{\pi/4,\,\epsilon,\,\epsilon^2,\,0\}$ and phases $0$ or $\pi$, where $x\in\{\ell,\nu\}$. We have truncated the series of $\theta_{ij}^x$ after $\epsilon^2$ and approximate $\epsilon^n=0$ for $n>2$, since this corresponds to the present experimental precision. In order to obtain the textures, we have expanded the mass matrices to second order in $\epsilon$ about $\epsilon=0$, and identified each element with its leading order in $\epsilon$. An obvious advantage of this method is that no diagonalization is needed. Out of the $262\,144$ obtained combinations, $2\,468$ are compatible with current experiments at the $3\sigma$ CL.
\\
\indent In \figu{histeps02}, we show the mixing angle distributions for these combinations. One can see that most of the cases could be tested in future experiments. The corresponding lepton mass matrices are obtained with \equ{lepdiag}. As an example of our results\footnote{Note, since usually more than one Yukawa coupling matrix leads to the same texture, we choose the one that fits experimental data best.}, we show in \Tab~\ref{tab:textures} textures \#17 and 18 of \cite{Plentinger:2006nb} with a perfect fit to tribimaximal mixing \cite{tbm} (the complete list including a notebook can be found in \cite{Plentinger:2006nb}). Thereby, we set $U_{\ell'}=\mathbbm{1}$ since it does not affect any observables.\footnote{However, note that $M_\ell$ depends, in contrast to $U_\mathrm{ PMNS}$, on $U_{\ell'}$.}
\\
\indent In \Tab~\ref{tab:textures}, we call the new texture of $M_\nu^\mathrm{ Maj}$ ``diamond'' texture because it has order one entries in the corners. A general feature of this kind of textures is that $\theta_{13}^\nu=\pi/4$. This texture is a direct result of our systematic approach. In addition, we have found new sum rules, \eg, for the textures in \Tab~\ref{tab:textures}, we obtain
\begin{equation}\label{equ:sumt12new}
\theta_{12}+\frac{3}{5+2\sqrt{2}}\,\epsilon =\arctan(2-\sqrt{2}).
\end{equation}
This should be compared with our result corresponding to the usual quark-lepton complementarity relation \cite{qlc}
\begin{equation}\label{equ:sumt12}
\theta_{12}+\frac{\epsilon}{\sqrt{2}}+\frac{\epsilon^2}{\sqrt{2}}=\frac{\pi}{4},
\end{equation}
which we have, of course, obtained as a special case from our procedure, as expected.

\section{Type-I Seesaw}
\label{sec:seesaw}
\begin{figure}[ht]
\begin{center}
\includegraphics[width=.34 \textwidth]{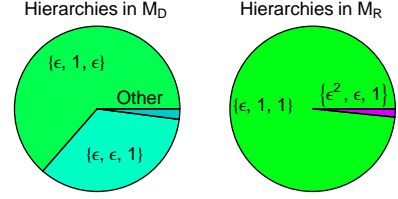}
\end{center}
\caption{\label{fig:mnorm} \footnotesize Distributions of hierarchies in $M_D$ (left) and $M_R$ (right)
leading to NH neutrino masses for all valid seesaw Yukawa coupling matrices (figure taken from \cite{Plentinger:2007px}).}
\end{figure}

\begin{figure}[ht]
\hspace{-1.cm}
\includegraphics[width=.6 \textwidth]{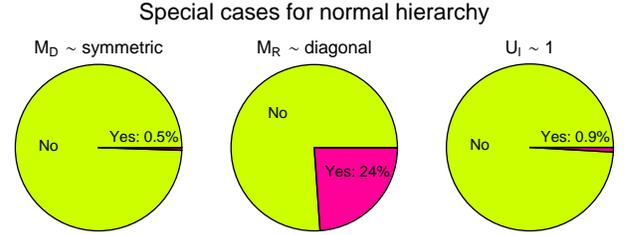}
\caption{\label{fig:special} \footnotesize The different pies show the fractions of special cases with symmetric $M_D$, diagonal $M_R$, and $U_\ell \simeq \mathbbm{1}$, of all allowed Yukawa coupling matrices (figure taken from \cite{Plentinger:2007px}). This figure is based on the mixing matrices, where $U_D \simeq U_{D'}$ in the first case, $U_R \simeq \mathbbm{1}$ and $U_\ell \simeq \mathbbm{1}$ in the second and third case, respectively. For the similarity condition ``$\simeq$'', we allow for $\epsilon^2$-deviations in the mixing angles. For instance, for an exact $U_R=\mathbbm{1}$, one would have only 2\% of all Yukawa coupling matrices.}
\end{figure}

Our procedure above can also be extended to the (type-I) seesaw mechanism \cite{seesaw}.\footnote{The $3\times3$ case in \Sec~\ref{sec:SM} can be directly realized in the type-II seesaw.} In that mechanism, the mass terms for the charged leptons, Dirac and Majorana neutrinos are
\be
\label{equ:seesawmassterms}
\begin{array}{cl}
 \mathcal{L}_\mathrm{ mass}=&-(M_\ell)_{ij}e_ie^c_j -(M_D)_{ij}\nu_i\nu_j^c-\\[.2cm]
&\frac{1}{2}(M_R)_{ij}\nu^c_i\nu^c_j+\mathrm{h.c.}
\end{array}
\ee
The mass matrices are diagonalized by
\be
\begin{array}{ll}
M_\ell=U_\ell M_\ell^\mathrm{ diag}U_{\ell'}^\dagger,\quad &M_D = U_D M_D^\mathrm{ diag} U_{D'}^\dagger,\\
M_R = U_R M_R^\mathrm{ diag} U_R^T,\quad &M_\mathrm{ eff}=U_\nu M_\mathrm{ eff}^\mathrm{ diag}U_\nu^T,
\end{array}
\label{equ:seesawmassmatrices}
\ee
where the effective neutrino mass matrix is given by the seesaw formula
\be
\label{equ:seesaw}
M_\mathrm{ eff}=-M_DM_R^{-1}M_D^T.
\ee
Together with the parameterization in \equ{unitary}, we obtain
\be
\label{equ:meffth}
\begin{array}{ll}
M^\mathrm{ th}_\mathrm{ eff}=&-D_D\widehat{U}_D\tilde{K}M_D^{\mathrm{diag}}\widehat{U}_{D'}^\dagger
\tilde{D}\widehat{U}_R^* (K_R^*)^2\times\\[.2cm]
& (M_R^{\mathrm{diag}})^{-1} \widehat{U}_R^\dagger \tilde{D}\widehat{U}_{D'}^*M_D^{\mathrm{diag}}\tilde{K}\widehat{U}_D^T D_D~,
\end{array}
\ee
where we have introduced $\tilde{K}=K_D^\ast K_{D'}$, and $\tilde{D}=D_{D'}^\ast D_R^\ast$. If we use, on the other hand, the experimentally known PMNS matrix in \equ{PMNS} together with \equ{PMNSmixings}, and insert it into \equ{seesawmassmatrices}, we obtain
\be
\label{equ:meffexp}
\begin{array}{ll}
M_\mathrm{ eff}^\mathrm{ exp}=& D_\ell\widehat{U}_\ell K_\ell\widehat{U}_\mathrm{PMNS}K^2_\mathrm{Maj}\times\\[.2cm]
& M^\mathrm{ diag}_\mathrm{ eff}\widehat{U}^T_\mathrm{PMNS}K_\ell\widehat{U}_\ell^TD_\ell~.
\end{array}
\ee
Since both, $M^\mathrm{ th}_\mathrm{ eff}$ and $M_\mathrm{ eff}^\mathrm{ exp}$ describe the same mass matrix, they are identical. Therefore, we match $M^\mathrm{ th}_\mathrm{ eff}$ and $M_\mathrm{ eff}^\mathrm{ exp}$ with a precision of $\epsilon^3$. In this way, we circumvent the diagonalization of $M_\mathrm{ eff}$.
For the resulting mass matrices, we show in \figu{mnorm} the distribution of their mass hierarchies, and in \figu{special} the fraction of special cases often considered in literature.
There, one can see that in most cases, $M_R$ has degenerate masses which could be used for resonant leptogenesis. Just a small fraction has a strictly hierarchical mass spectrum in $M_R$ which would be preferred in usual leptogenesis scenarios. Special cases, such as symmetric $M_D$ or diagonal $M_\ell$, are less than $1\%$, contrary to diagonal $M_R$, which represents $24\%$ of all cases. As an example, we show in \Tab~\ref{tab:seesawtextures} the seesaw textures \#17 and 18 of \cite{Plentinger:2007px} with the corresponding Yukawa coupling matrices. Note that $\theta_C$ and the mass ratios are hardly affected by renormalization group effects since the correction to the lepton mixing is $\lesssim 1^\circ$ for NH neutrinos.
\begin{table*}
\begin{center}
\begin{tabular}{|@{\hspace{1mm}}c@{\hspace{1mm}}||@{\hspace{0.5mm}}c@{\hspace{0.5mm}}|@{\hspace{0.5mm}}c@{\hspace{0.5mm}}|@{\hspace{0.5mm}}c@{\hspace{0.5mm}}||@{\hspace{0.5mm}}c@{\hspace{0.5mm}}|@{\hspace{0.5mm}}c@{\hspace{0.5mm}}|@{\hspace{0.3mm}}c@{\hspace{0.3mm}}||@{\hspace{0.3mm}}c@{\hspace{0.3mm}}|}
\hline &&&&&&& \vspace{-3mm} \\*
\# & $M_\ell$ & $M_D$ & $M_R$ & 
$\begin{array}{@{\hspace{0.5mm}}c@{\hspace{0.5mm}}}
 {m_i^D}/{m_D} \\[1mm]
 {m_i^R}/{M_\mathrm{ B-L}}
\end{array}$
 & 
$\begin{array}{@{\hspace{0.5mm}}c@{\hspace{0.5mm}}}
 (\theta _{12}^\ell,  \theta _{13}^\ell, \theta _{23}^\ell) \\[1mm]
 (\theta _{12}^D, \theta _{13}^D, \theta _{23}^D) \\[1mm]
 (\theta _{12}^{D'}, \theta _{13}^{D'},  \theta _{23}^{D'}) \\[1mm]
 (\theta _{12}^R, \theta _{13}^R,  \theta _{23}^R) 
\end{array}$
& \hspace{-.3cm}\begin{minipage}{2.5cm}\begin{tabular}{c} $( \delta ^l \hspace{-0.5mm}, \alpha _1^l , \alpha _2^l ) $ \\
$( \delta ^D \hspace{-0.5mm}, \varphi _1^D , \varphi _2^D , \varphi _3^ D ) $ \\
$( \delta ^{D'} \hspace{-1.5mm}, \alpha _1^{D'} , \alpha _2^{D'} ) $ \\
$( \delta ^R \hspace{-0.5mm}, \varphi _1^R , \varphi _2^R , \varphi _3^R ) $\end{tabular}\end{minipage}&
$( \theta_{12} , \theta_{13} , \theta_{23} ) $ 
\\&&&&&&&\vspace{-3.4mm}
\\ \hline&&&&&&&\vspace{-3mm}\\ $17$ & $
\left(
\begin{array}{ccc}
 0 & \epsilon ^2 & 1 \\
 0 & \epsilon ^2 & \epsilon  \\
 0 & \epsilon ^2 & 1
\end{array}
\right)$ & $\left(
\begin{array}{ccc}
 \epsilon  & \epsilon ^2 & \epsilon  \\
 1 & \epsilon  & 1 \\
 1 & \epsilon  & 1
\end{array}
\right)$ & $\left(
\begin{array}{ccc}
 1 & \epsilon  & 1 \\
 \epsilon  & 1 & \epsilon  \\
 1 & \epsilon  & 1
\end{array}
\right)$ & $
\begin{array}{c}
 ( \epsilon  , \epsilon  , 1 ) \vspace{0.1cm}\\
 ( \epsilon  , 1 , 1 )
\end{array}
$ & $
\begin{array}{c}
 ( \frac{\pi }{4} , \frac{\pi }{4} , \epsilon  ) \\
 ( \epsilon  , \epsilon^2 , \frac{\pi }{4} ) \\
 ( 0 , \frac{\pi }{4} , \epsilon^2 ) \\
 ( \epsilon  , \frac{\pi }{4} , \epsilon^2 )
\end{array}
$& \hspace{.4cm}\begin{minipage}{2.5cm}\begin{tabular}{c}
$
\begin{array}{c}
(\pi ,0,0)
\end{array}
$ \\ $
\begin{array}{c}
(\pi ,0,0,0)
\end{array}
$ \\ $
\begin{array}{c}
(0,0,\pi )
\end{array}
$ \\ $
\begin{array}{c}
(\pi ,0,0,\pi )
\end{array}$
\end{tabular} \end{minipage}& $
\begin{array}{c}
(33.3^\circ,0.0^\circ,51.4^\circ)
\end{array}
$
\\&&&&&&&\vspace{-3.4mm}
\\ \hline&&&&&&&\vspace{-3.2mm}\\ $18$ & $
\left(
\begin{array}{ccc}
 0 & \epsilon ^2 & 1 \\
 0 & \epsilon ^2 & 1 \\
 0 & \epsilon ^2 & 1
\end{array}
\right)$ & $\left(
\begin{array}{ccc}
 \epsilon  & \epsilon  & 0 \\
 \epsilon  & \epsilon  & \epsilon  \\
 \epsilon ^2 & \epsilon ^2 & 1
\end{array}
\right)$ & $\left(
\begin{array}{ccc}
 1 & 1 & 0 \\
 1 & 1 & 0 \\
 0 & 0 & 1
\end{array}
\right)$ & $
\begin{array}{c}
 ( \epsilon  , \epsilon  , 1 ) \vspace{0.1cm}\\
 ( \epsilon  , 1 , 1 )
\end{array}
$ & $
\begin{array}{c}
 ( \frac{\pi }{4} , \frac{\pi }{4} , \frac{\pi }{4} ) \\
 ( \frac{\pi }{4} , 0 , \epsilon  ) \\
 ( \epsilon^2 , \epsilon^2 , \epsilon^2 ) \\
 ( \frac{\pi }{4} , \epsilon  , \epsilon  )
\end{array}
$& \hspace{.4cm}\begin{minipage}{2.5cm}
\begin{tabular}{c}
$\begin{array}{c}
(0,\pi ,\pi )
\end{array}
$ \\ $
\begin{array}{c}
(0,0,\pi ,\pi )
\end{array}
$ \\ $
\begin{array}{c}
(0,\pi ,\pi )
\end{array}
$ \\ $
\begin{array}{c}
(0,0,\pi ,\pi )
\end{array}
$\end{tabular}\end{minipage} & $
\begin{array}{c}
(33.5^\circ,0.2^\circ,51.4^\circ)
\end{array}$
\\[6.1mm] \hline
\multicolumn{8}{c}{}\vspace{-.15cm}\\
\end{tabular}
\end{center}
\caption{Selected examples of seesaw textures and Yukawa coupling matrices from \cite{Plentinger:2007px}.}
\label{tab:seesawtextures}
\end{table*}
\\
\indent Such a comprehensive list of mass matrix textures and Yukawa coupling matrices can be used to explore the huge parameter space as well as for model building. As a proof of principle, we assume a 7-fold $Z_4$ product flavor symmetry $G_F$ with two SM singlet flavons $f_i$ and $f'_i$ per $Z_4^i$, that are charged as $f_i\sim 1$ and $f'_i\sim 2$, and have universal VEVs $\langle f_i\rangle\simeq\langle f_i'\rangle\simeq v$, for $i=1,2,\dots, 7$. This flavor symmetry generates via the Froggatt-Nielsen mechanism \cite{Froggatt:1978nt} and the charge assignment in \Tab~\ref{tab:ZNcharges} the mass matrix textures \#17 and 18 of \Tab~\ref{tab:seesawtextures} (\cf~also \cite{Plentinger:2007px}), respectively.
\begin{table}
\begin{center}
\begin{tabular}{c||c||c}
\hline
Field & Model 1& Model 2\\
\hline 
$\nu^c_1$ & (0,0,0,1,0,1,1) &  (2,0,0,2,0,0,1) \\ 
$\nu^c_2$ & (2,0,0,1,0,1,1) &  (2,0,0,2,0,0,1) \\ 
$\nu^c_3$ & (0,0,0,1,0,1,1) &  (0,2,0,0,2,1,0)\\
\hline 
$\ell_1$ & (0,2,0,0,1,0,1) &  (2,0,2,2,2,1,0) \\ 
$\ell_2$ & (0,0,0,0,1,1,0) &  (2,2,0,2,2,1,0)\\ 
$\ell_3$ & (0,0,2,1,0,0,1) &  (0,2,2,2,2,0,1) \\
\hline 
$e^c_1$ &  (2,2,2,1,1,1,1) & (0,0,0,0,0,1,1)\\ 
$e^c_2$ &  (2,2,2,0,0,1,0) & (2,2,2,0,0,3,3)\\ 
$e^c_3$ &  (0,2,2,0,0,0,0) & (2,2,2,2,2,3,3)\\ 
\hline
\end{tabular}
\end{center}
\label{tab:ZNcharges}
\caption{\footnotesize Assignment of flavor charges to the leptons. The models 1 and 2 lead respectively to the textures \#17 and 18 in \Tab~\ref{tab:seesawtextures} (table taken from \cite{Plentinger:2007px}).}
\end{table}
Since these textures are contained in our list, they are guaranteed to provide a realistic fit to the lepton masses and mixings. The necessary order one Yukawa couplings can be easily reconstructed by using the given parameters. Note, since we set $U_{\ell'}=\mathbbm{1}$, we list just a fraction of the viable textures. That means that the introduction of non-trivial $U_{\ell'}$ increases the number of valid textures which is useful for a systematic identification of flavor symmetries.

\section{Outlook}
\label{sec:outlook}
In this talk, we have presented a comprehensive list of textures for effective lepton masses and for the seesaw mechanism. The mass matrices are obtained via the hypothesis of extended quark-lepton complementarity, \ie, all masses and mixings are assumed to be powers of $\epsilon\simeq\mathcal{O}(\theta_C)$. For the mixings, the zeroth order in $\epsilon$ is interpreted as a maximal mixing angle $\pi/4$.
\\
\indent As a direct consequence of our systematic approach, we have obtained at least one new texture and a new sum rule. In addition, we have shown the distribution of PMNS mixing angles and how they could be tested by future experiments. We have also shown that only a small fraction of our results would favor usual leptogenesis scenarios, while most of the cases could be used for resonant leptogenesis. Also special cases of symmetric $M_D$ and $M_\ell$ are just a small fraction of less than $1\%$, whereas diagonal $M_R$ represents 24\% of all resulting cases. A more sophisticated analysis can be found in \cite{Plentinger:2006nb,Plentinger:2007px}. This includes details of our procedure, the complete list of textures, the discussion of inverted hierarchical and degenerate neutrinos, variations of $\epsilon$ and Dirac phases, distributions for $0\nu\beta\beta$ \etc
\\
\indent We have also shown how our textures can be generated in explicit models from flavor symmetries. This could be regarded as proof of principle that our approach can be used both to explore the parameter space in a possibly less biased way as well as for model building. Thereby, a systematic identification of flavor symmetries for textures could be done in an ``automated'' way.

\subsubsection*{Acknowledgments}
I would like to thank my collaborators G. Seidl and W. Winter. The research of F.P. is supported by Research Training Group 1147 \textit{Theoretical Astrophysics and Particle Physics} of Deutsche Forschungsgemeinschaft.

\end{document}